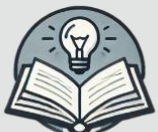

# The Patentist Living Literature Review

# International trade and intellectual property

Gaétan de Rassenfosse

Holder of the Chair of Science, Technology, and Innovation Policy
École polytechnique fédérale de Lausanne, Switzerland.

This version: June 2025

**Purpose**
This article is part of a Living Literature Review exploring topics related to intellectual property, focusing on insights from the economic literature. Our aim is to provide a clear and non-technical introduction to patent rights, making them accessible to graduate students, legal scholars and practitioners, policymakers, and anyone curious about the subject.

**Funding**
This project is made possible through a Living Literature Review grant generously provided by Open Philanthropy. Open Philanthropy does not exert editorial control over this work, and the views expressed here do not necessarily reflect those of Open Philanthropy.

# International Trade and Intellectual Property

Gaétan de Rassenfosse
École polytechnique fédérale de Lausanne, Switzerland.

International trade rarely leaves the headlines. Politicians fret over bilateral deficits, pundits debate tariff hikes and supply-chain "re-shoring," while analysts track shipping costs and trade volumes. Yet trade costs, including tariffs, are only part of the story. Trade flows also pivot on structural fundamentals—geography and distance, relative factor endowments, technological capabilities, consumer preferences, and firms' competitiveness—as well as on historical and cultural ties such as common languages or colonial legacies. Overlaying these fundamentals is an often-invisible architecture of rules and institutions that determine who can sell what to whom, and on what terms.

A central piece of that architecture is intellectual property (IP) protection. The rise of knowledge-intensive manufactures—from biotech reagents and advanced semiconductors to design-rich consumer electronics—and the explosive growth of digitally delivered products and services have made legal rights governing copying, licensing, and data access critical determinants of market access. Since the World Trade Organization's 1995 Agreement on Trade-Related Aspects of Intellectual Property Rights ([TRIPS](#)), and its "TRIPS-plus" upgrades in mega-regional agreements, IP provisions have moved from the margins of trade negotiations to their core. But how exactly does IP affect global commerce?

**Countervailing forces**

The overall theoretical framework for studying the effect of intellectual property rights (IPRs) on international trade primarily revolves around two countervailing forces, as explained by Maskus and Penubarti (1995).

A first force is the *market power effect*, which lowers the quantity exported. Stronger IPRs mean that it is harder for rivals in the destination market to imitate the technology sold by the foreign firm. In economic terms, the supply of close substitutes shrinks, and the foreign firm can exercise increased monopolistic power. This leads the firm to charge a higher price and ship fewer units, lowering export volume even though per-unit profits rise.

A second, opposing force is the *market expansion effect*, which boosts exports. Tighter enforcement lowers the risk that local imitators will [free-ride](#) on the exporter's technology or brand. This increases the demand for the original products in several ways: by limiting the number of substitute products available and by increasing the value of the original product. With imitation less likely, the producer finds it worthwhile to offer robust warranties, after-sales service, and complementary inputs (such as software updates and spare parts), thereby raising customers' willingness to pay. In economic terms, this result is an outward shift of the [residual demand curve](#)—at every possible price, the exporter can sell a larger quantity.

The net impact on trade flows depends on which force dominates, but some broad theoretical generalizations can be made. For example, the market power effect is likely to be most visible in situations with strong parallel-trade restrictions, so arbitrage—where products sold at lower prices in one market are resold in another where prices are higher—cannot undo the price increase. The market expansion effect may dominate in sectors where brand integrity and after-sales service are crucial, such as the medical device and luxury apparel industries.

The effects described above are "static," in the sense that they consider how strengthening IPRs affects the existing set of products. However, there are also "dynamic" effects, which lead firms to improve their products (quality upgrading) or introduce new ones (product differentiation). The mechanism works as follows. Stronger IPRs reduce the risk of imitation by domestic firms in importing countries, which in turn increases the expected returns on innovations. This increased security incentivizes exporting firms to invest more in R&D to create new products and improve existing ones. With better protection for their intellectual assets, firms are more willing to bring differentiated and higher-quality products to market. This can involve climbing the quality ladder (vertical differentiation) or creating new varieties that set them apart from rivals (horizontal differentiation).

**Measuring the effect is challenging**

Quantifying the effect of IP on international trade is no small matter, for scholars must address several empirical challenges. First, measuring the level of IP protection is notoriously difficult. We have discussed in a previous article the various indices that have been proposed, capturing both the law on the books and the law in practice.[1] These indices, despite providing significant structure, may still lack the nuances necessary to fully grasp the various facets of the IP regime that affect trade. Many indices focus solely on patents, overlooking other forms of IPRs, such as trademarks, copyrights, and trade secrets. There is also a lack of comprehensive, time-varying data on certain IP-related measures (*e.g.*, parallel trade, pricing regulations), making fine-grained analyses challenging.

Second, and at least as challenging, is establishing causality. Do stronger IPRs cause trade increases, or do expanding trade flows and rising economic development (or the prospect thereof) persuade governments to tighten IP protection? Trade flows are shaped by a host of hard-to-observe forces (*e.g.*, macroeconomic reforms, institutional quality, geopolitical risks) that often correlate with changes in IPRs, and teasing out the independent effect of IP on international trade is difficult. The problem is compounded by policy bundling: IP upgrades rarely arrive in isolation. Modern "deep" trade agreements bundle IP chapters with liberalisation of services, investment, and public procurement. It is thus challenging to separate the effect of IP provisions from other trade-related issues within the same trade agreements. Unless researchers can disentangle these concurrent reforms, any econometric coefficient labelled "IP strength" is liable to absorb the impact of the whole package rather than the IP provisions themselves.

**An evolving empirical literature**

Despite these hurdles, a sizeable body of empirical work has emerged over the past three decades that quantifies the impact of IP on trade. These studies differ in several ways, resulting in distinct strands of evidence. A first dividing line runs between studies that examine general shifts in IP strength—typically proxied by the Ginarte-Park index—and those that exploit specific reforms, such as TRIPS. The early literature primarily belongs to the former camp, where causal claims are more difficult to sustain for the reasons outlined above. More recent research treats policy events like TRIPS as quasi-natural experiments: by inducing many developing countries to tighten their IP regimes, the agreement provides an exogenous shock that helps isolate the effect of stronger IP from other confounding factors. A newer strand goes further, matching trade flows to the actual patenting behaviour of exporting firms, thereby complementing country-level IP indicators with actual use of IP by firms.

---

[1] de Rassenfosse, G. (2025). The strength of patent systems. *The Patentist Living Literature Review* 5: 1–5. DOI: 10.48550/arXiv.2505.07121.

A second line of demarcation is the unit of analysis. Work has progressed from aggregate trade flows to increasingly granular data—industry-, firm-, and even product-class level. Greater disaggregation reveals how IP effects differ across sectors (high-tech vs. low-tech, pharmaceuticals vs. agriculture) and across destinations with varying imitation risk. It also allows researchers to unpack the mechanisms at play, distinguishing, for example, between changes at the extensive margin (new product varieties or quality upgrades) and the intensive margin (export volumes of existing products).

**Yes, IP matters. But how depends on context**

Early studies using aggregate data yielded mixed results, sometimes positive, negative, or insignificant, depending on factors such as the level of patent protection, country types, or goods traded. Some seminal works, like Maskus and Penubarti (1995), have found that stronger patent protection generally has a positive impact on bilateral manufacturing imports, particularly in developing economies, indicating a market expansion effect. Smith (1999) introduced the concept that the impact depends on the importing country's imitative ability: market expansion effects dominated for countries with high imitation threat, while market power effects were more prevalent for those with weak imitation threat.

With the implementation of TRIPS, studies increasingly treated this as an exogenous policy shock, allowing for stronger causal inferences. Research has found that the implementation of TRIPS is associated with increased trade in knowledge-intensive goods, particularly exports by developed countries to developing countries, and in specific sectors such as ICT and biopharmaceuticals (Delgado et al. 2013). Ivus (2010) similarly found that stronger patent rights in developing countries, due to TRIPS, increased developed countries' exports, especially in patent-sensitive industries such as medicinal and pharmaceutical products. This increase was driven by an expansion in the quantity of exports, rather than higher prices, suggesting that it did not limit access to innovative products.

The more recent literature disaggregates the quantity effect along the *extensive margin* (the variety of goods) and the *intensive margin* (the volume of existing goods). There appears to be a consensus that stronger IPRs primarily boost trade through the *extensive margin*, by increasing the variety of products traded, rather than the *intensive margin* (*e.g.*, Ivus, 2015; Martínez-Zarzoso and Santacreu, 2025). This implies that IPRs encourage firms to introduce new products into foreign markets.

**Beyond shipments at the border**

The effect of IP on international trade also manifests through different channels that may take a longer time to materialize but can have a significant impact on trade flows. Stronger IPRs can influence a firm's choice of how to serve a foreign market, potentially shifting from exporting to licensing or [foreign direct investments](#) (FDI), as discussed by Branstetter et al. (2006).

If patents and trade secrets are well enforced in the destination country, the foreign innovator may choose to license its know-how to a local producer rather than exporting the finished good. Exports of physical goods can fall while cross-border royalty flows rise. Similarly, stronger IP protection lowers the risk that proprietary know-how leaks out of a foreign factory. Firms are, therefore, more willing to build or buy production facilities abroad, especially for R&D-intensive or design-heavy stages of the supply chain. Trade in intermediate parts (*i.e.*, components sent to the affiliate) can increase, while finished-goods exports can rise if final assembly now takes place locally.

Put simply, strong IP can persuade a company to "make or license there" rather than "ship from here." Customs data alone might show smaller export volumes, but that needn't signal lost business: the economic activity has merely shifted into royalties, intra-firm trade, and local production.

**Looking ahead**

Digital and other remotely-deliverable services are a growing component of global trade. Unlike containerized goods, these flows depend almost entirely on a web of intangible rights, including copyrights in software and creative content, patents on algorithms, database rights, and trade-secret protection for proprietary code and know-how. Yet the empirical base lags far behind their economic weight. Streaming royalties collected through app stores, SaaS subscriptions paid to cloud providers, in-app micro-payments, and cross-border data-processing fees rarely appear—at least not cleanly—in customs or balance-of-payments statistics. Until those statistical blind spots are filled, our understanding of how IPRs shape twenty-first-century trade will remain fragmentary. Strengthening measurement, therefore, appears as one of the most pressing tasks for researchers and policymakers seeking evidence-based guidance on the next generation of IP and digital trade agreements.